\documentclass[prd,superscriptaddress,amsfonts,amssymb,amsmath,showpacs,twocolumn]{revtex4-2}
\usepackage{bm}
\usepackage{amsfonts}
\usepackage{latexsym}
\usepackage{graphicx}
\usepackage{amsmath}
\usepackage{palatino}
\usepackage{mathpazo}
\usepackage{textcomp}
\usepackage{float}
\usepackage{booktabs}
\usepackage{dcolumn}
\usepackage{booktabs}
\usepackage{multirow}
\usepackage{hyperref}
\hypersetup{colorlinks,citecolor=blue}
\usepackage{amsmath}
\usepackage{xcolor}
\usepackage{orcidlink}
\usepackage{epsfig}
\usepackage{caption}
\usepackage{subcaption}
\usepackage{commath}
\hypersetup{colorlinks,citecolor=blue}
\hypersetup{colorlinks=true,linkcolor=magenta,filecolor=magenta,urlcolor=blue}

\usepackage{amssymb}
\usepackage{relsize,exscale}
\DeclareUnicodeCharacter{2265}{~}

\def\be{\begin{equation}}
\def\ee{\end{equation}}
\def\bea{\begin{eqnarray}}
\def\eea{\end{eqnarray}}

\begin{document}

\title{\bf How parameter constraining can influence the mass accretion process of a Black Hole in the Generalized Rastall Gravity Theory ?}
\author{Puja Mukherjee}
\email{pmukherjee967@gmail.com} 
\affiliation{Department of
Mathematics, Indian Institute of Engineering Science and
Technology, Shibpur, Howrah-711 103, India}
\author{Ujjal Debnath}
\email{ujjaldebnath@gmail.com} 
\affiliation{Department of
Mathematics, Indian Institute of Engineering Science and
Technology, Shibpur, Howrah-711 103, India}
\author{Himanshu Chaudhary}
\email{himanshu.chaudhary@ubbcluj.ro,\\
himanshuch1729@gmail.com}
\affiliation{Department of Physics, Babeș-Bolyai University, Kogălniceanu Street, Cluj-Napoca, 400084, Romania}
\affiliation{Research Center of Astrophysics and Cosmology, Khazar University, Baku, 41 Mehseti Street, AZ1096, Azerbaijan}
\author{G. Mustafa}
\email{gmustafa3828@gmail.com}
\affiliation{Department of Physics,
Zhejiang Normal University, Jinhua 321004, People’s Republic of China}


\begin{abstract}\label{A}
\noindent
Black holes, one of the greatest enigmas of our Universe, are challenging to decipher. This work is dedicated to observing the changes in the mass of a non-singular black hole with the evolution of the Universe in the generalized Rastall gravity framework, considering the effects of parameter constraining. We examine two recently developed dynamical dark-energy equation of state parameterization models: Chaudhary-Bouali-Debnath-Roy-Mustafa-type~(CBDRM)~parameterization and Chaudhary-Debnath-Mustafa-Maurya-Atamurotov-type~(CDMMA)~parameterization. Starting with the concept and fundamental equations of the generalized Rastall gravity theory, we introduce the two models along with their equations of state, energy density equations, and corresponding Hubble parameter equations. We then constrain the required parameters using Monte Carlo Markov chain (MCMC) analyses to ensure the accuracy and reliability of our study. Next, we discuss the non-singular black holes from the perspective of generalized Rastall gravity theory and some of their essential properties. Finally, we pursue the primary goal of our work: analyzing the mass accretion process. We derive the mass equation for both models in terms of the redshift function, represent the results graphically, and compare them with the standard $\Lambda$CDM model of the Universe. Our findings indicate that the accretion of both CBDRM~and~CDMMA dark energy parameterizations, considering constrained parameter values, leads to an increase in the mass of the black hole during the Universe's evolution within the generalized Rastall gravity theory, revealing the true nature of dark energy.
\end{abstract}


\maketitle
\textbf{Keywords:}~Dark Energy Parametrization~,~Black Hole~,~Generalized Rastall Gravity~,~Posterior Inference~,~Mass Accretion~.
\section{Introduction}\label{In}
\textbf{``For myself, I like a Universe that includes much that is unknown and, at the same time, much that is knowable"} -this famous quote by Carl Sagan shows how this vast Universe contains so many mysteries in itself but is always ready to be discovered; all its need is more and more exploration by us. That is why Sir Carl Sagan again mentioned in Cosmos, \textbf{``We are a way for the cosmos to know itself.”} Currently, our Universe is in its accelerated expansion phase \cite{perlmutter1998cosmology,bahcall1999cosmic,filippenko1998results,verde20022df}. The most acknowledged candidate responsible for this type of behavior in the Universe is a peculiar kind of energy that exists almost everywhere in the Universe and has a positive energy density but negative pressure. For its mysterious appearance, it is called dark energy. Over time cosmologists developed many forms of dark energy, be it static like the infamous cosmological constant \cite{padmanabhan2003cosmological,carroll1992cosmological} or dynamical dark energies \cite{peebles1988cosmology,caldwell2002phantom,armendariz2000dynamical,gasperini2001quintessence,sen2002rolling,wei2005hessence,martin2008dbi,li2004model}. Also, considering quantum gravity theory, other prospective dark energy models were created \cite{cai2007dark,hao2008cosmological}. However, any kind of dark energy model is prone to some issues related to misinterpreting the original nature of dark energy and the divergence problem. As a solution, we can effectively use the parameterized form of the dark energy equation of state as it is a convenient way for us to study the expansion of the universe \cite{sardar2024cosmography}. This particular approach to considering dark energy parameterization consists of many key advantages such as being model-independent, which allows us to explore the properties of dark energy in a more general and flexible manner without any limitation of a particular theoretical framework. Parameterizations are often drafted to be mathematically compliant and simple so they can easily fit into observational datasets, be it from cosmic microwave background (CMB)~,~supernovae~,~or baryon acoustic oscillations (BAO)~, making it easier for us to constrain the properties of various dark energies. Again, this process of parameterization is free of theoretical biases, unlike some specific dark energy models like quintessence, phantom energy, or Chaplygin gas, leading to a more reliable and data-driven method to understand dark energy. This process of parameterizations of the dark energy equation of state can very well accommodate a wide range of dark energy behaviors, such as the constant equation of state~($\omega=constant$)~or evolving equation of state~($\omega(z)$) making them suitable for testing deviations from the usual cosmological constant without considering any specific model. It normally involves very few free parameters, resulting in less computational complexity and more simplified statistical analysis than any other complex model with many degrees of freedom. By analyzing any common parameterization model, we can consistently compare different observational datasets and theoretical models, drawing a conclusion about their consistency with each other. Some parameterization models allow us to observe features of time-dependent dark energies, which is crucial for distinguishing the differences between the usual cosmological constant and dynamical dark energy models. Unlike dark energy models involving too many parameters which can lead to overfitting of observational data, parameterizations create a balance by using a minimal set of parameters to describe the equation of state. Most importantly, with such a rapid development in observational cosmology, whenever new data becomes available, we can update and extend parameterizations at our own convenience to incorporate new insights without completely remodeling our previous theoretical frameworks. Even sometimes parameterizations are also used in forecasting studies about the ability of future experiments like  Euclid, LSST, or DESI to constrain the properties of dark energy. So, although there are some limitations of this process regarding the choice of parameterization and its extrapolation beyond the observed redshift range which can lead to some unphysical results as well as this process may not be able to capture all the possible behaviors of dark energy specifically when the actual equation of state has complex features. Still, this parameterization process provides a more flexible and data-driven approach to studying dark energies, making this a potent tool of modern cosmology as it is often complemented by some specific theoretical models, giving it deeper physical insights. That is why of late, some recently developed dark energy parameterizations were studied in \cite{debnath2024dark,khurana2024exploring,chaudhary2024addressing}~,~opening a new door for researchers to explore.\\
In this paper, we take advantage of this opportunity to study two of these recently developed dynamical parameterization models of the dark energy equation of state, namely Chaudhary-Bouali-Debnath-Roy-Mustafa-type (CBDRM) parameterization~and~Chaudhary-Debnath-Mustafa-Maurya-Atamurotov-type (CDMMA) parameterization. To study the effects of these dark energy parameterizations, we consider a very enigmatic object of nature known as a black hole. To describe a black hole, we can say \textbf{``There is an inexorable force in the cosmos, a place where time and space converge"} or \textbf{``Man has reached the most mysterious and awesome corner of the Universe... a point where here and now become forever"}, both of this tagline from the famous film ``The Black Hole~(1979)" can describe the concept of this peculiar thing to a great extent to any laymen. Now, what better way to study the influence of these dark energy parameterizations on black holes than by using their mass accretion process, as mass is the most important physical property of a black hole? Numerous works have been accomplished on the mass accretion phenomena of black holes. First, Bondi introduced this concept \cite{bondi1952spherically}, and then Michel improvised it in his work \cite{michel1972accretion}. After that, cosmologists took a keen interest in this topic, which led to a lot of work on this particular topic in the context of different types of black holes
\cite{babichev2004black,babichev2005accretion,jamil2009evolution,sharif2012phantom,debnath2015accretion,mukherjee2023accretion,dutta2019dark,debnath2015accretions}.\\\\
According to cosmologists, there is another feasible explanation for the accelerated expansion of the Universe: modified gravity theory. Several researchers studied the black hole mass accretion process in different gravity scenarios \cite{abbas2013thermodynamics,dwivedee2014evolution,mukherjee2024accretionphenomenadifferentkinds}. In addition, one of the prospective candidates for modified gravity theory is Rastall gravity, a more generalized form of Einstein's general theory of relativity \cite{rastall1972generalization,darabi2018einstein,heydarzade2017black,lobo2018thermodynamics,kumar2018rotating,mondal2022thermodynamics}. However, some studies still exhibit the challenges of explaining the accelerated expansion of the Universe through the usual Rastall gravity theory \cite{capone2009accelerating,batista2010testing}. So, the need of the hour becomes a further generalization of this concept \cite{moradpour2017generalization,lin2020cosmic}~,~giving rise to a new concept of gravity called generalized Rastall gravity. Various perspectives on this particular theory of gravity were already observed \cite{ziaie2020structure,das2018cosmological,mota2022generalized,moradpour2019black,lin2019neutral}. In this work, we will explore the mass accretion process of a non-singular black hole in the context of generalized Rastall gravity theory considering the effects of parameter constraining on it. In recent times, constraining parameters gained significant importance in cosmology because of their properties of improvement and removal of degeneracies in the data. Mainly, parameter constraints can improvise the equation of state parameter of dark energy~,~leading to more precise results. Quite a lot of research was conducted using the parameter constraining process in different gravity scenarios
\cite{lu2008constraints,xu2012modified,mukherjee2024constraining,chaudhary2023constraints,khurana2024exploring}. As mentioned earlier we choose to study two of the emerging parametrized form of dark energy equation of state,~~CBDRM~and~CDMMA~as these two involves a specific mathematical framework of theoretical physics mainly concerning black hole solutions, modified gravity theories, and cosmological models, exactly what is needed for our current study. Both of these parameterizations are designed to simplify complex physical systems, making them more compliant for analysis while preserving their essential physical features. They can easily incorporate generalizations of known solutions like Schwarzschild or Kerr black holes to more complex scenarios such as black holes in modified gravity or with some additional fields (e.g., electromagnetic or scalar fields). They have a clear physical interpretation and are flexible enough to accommodate various physical scenarios, making them useful for studying a wide range of phenomena, from the black hole mass accretion process to black hole thermodynamics to cosmological evolution. Also, in the context of modified gravity theory~,~these parameterizations are quite helpful in bridging the gap between observational phenomena and physical predictions, providing us the much-needed insights into how deviations from general relativity can be manifested in physical systems. In the framework of generalized Rastall gravity theory~,~these two parameterizations have significant physical relevance as they provide us with a systematic way to study the implications of the non-conserved stress-energy tensor and the matter-geometry coupling. They can very well be helpful to us in exploring black hole solutions~,~cosmological models~,~astrophysical systems~,~and thermodynamic properties in a way that is consistent with the principles of generalized Rastall gravity theory. By simplifying the field equations and incorporating observational constraints~,~these parameterizations can help to test and constrain generalized Rastall gravity~,~offering insights into its validity as an alternative to general relativity. So, these parameterizations are the perfect candidates to study the mass accretion process of non-singular black holes in the environment of generalized Rastall gravity theory.\\
All of the above references give us the required motivation to study the influence of parameter constraining on the mass accretion process of a non-singular black hole in generalized Rastall gravity theory. The structure of this paper is arranged in the following manner~: In Section \ref{S1}~,~we discuss the concept of generalized Rastall gravity theory and some basic equations related to it in detail. In addition, we introduce two types of dark energy equation of state parameterization methods as \textbf{Model-1}~(CBDRM)~and~\textbf{Model-2}~(CDMMA)~one by one with their corresponding equations of state~,~energy densities~,~and Hubble parameter equations~,~respectively. Then, in Section \ref{S2}~,~we explain the methodology and the description of the data. Here~,~we constrain the required parameters using the Monte Carlo Markov chain~(MCMC) technique and briefly compare the two considered models in the context of observations. Next, we discuss the non-singular black holes~,~their properties~, and the mass accretion process in the generalized Rastall gravity scenario in Section \ref{S3}. We first establish the results of the mass accretion process of both models through mass vs. redshift graphs and then compare them graphically with the standard $\Lambda$CDM model of the Universe. Finally, we provide a detailed discussion of the results found in this work and a conclusion with the necessary remarks in Section \ref{S4}.\\   
\section{The Concept and Basic Equations of Generalized Rastall Gravity}\label{S1}
Rastall questioned the usual law of energy-momentum conservation in curved spacetime and established a new concept that the rate of energy-momentum transfer between geometry and matter fields is not affected in any way by the evolution of the Universe \cite{moradpour2017generalization}. This consideration indicates that the coupling coefficient between the Rastall parameter, which is nothing but the derivative of the Ricci scalar, and the divergence of the energy-momentum tensor is constant, popularly known as the Rastall theory of gravity \cite{rastall1972generalization}. It is an example of modified gravity theory, in which a known divergence-free energy-momentum tensor is contemplated. However, this theory favors the effect of dark energy on the accelerated expansion of the Universe \cite{majernik2003cosmological,arbab2003cosmological,singh2020aspects,batista2012rastall}. Now, considering a more general approach where the Rastall parameter is no longer a constant gives rise to a new modified gravity theory called generalized Rastall theory. In this theory, a further generalized form for the divergence of the energy-momentum tensor is given as follows \cite{ziaie2020structure,das2018cosmological,moradpour2017generalization}:
\begin{equation}\label{1}
T^{\mu\nu}_{;\mu}=(\lambda R)^{,\nu}~,   
\end{equation}
Where $\lambda$ symbolizes the varying Rastall parameter given as \cite{moradpour2017generalization}:
\begin{equation}\label{2}
\lambda=\frac{\delta}{R}~,    
\end{equation}
$R$ and $\delta$ are the usual Ricci scalar and an unknown parameter. Now, following the Bianchi Identity, Einstein's equation takes the following form:
\begin{equation}\label{3}
G_{\mu\nu}+\epsilon\lambda g_{\mu\nu}=\epsilon T_{\mu\nu}~,    
\end{equation}
with $\epsilon$ being the Rastall gravitational constant \cite{moradpour2017generalization}. Eqn.(\ref{1}) reduce to the original Rastall theory \cite{rastall1972generalization} whenever $\lambda=$~constant. Again, to obtain the original Einstein field equations from the modified field equations mentioned above~,~we need to take the limit $\lambda \to 0$, which means that geometry and matter field are coupled to each other only in the simplest way. Alternatively, the Rastall parameter $\lambda$ used in Eqn.(\ref{3}) can be considered nothing but a varying cosmological constant in Einstein's gravity. Considering the line element of a spatially flat~,~isotropic and homogeneous Universe given by:
\begin{equation}\label{4}
ds^2=-dt^2+a^2(t)~[dr^2+r^2~(d\theta^2+sin^2\theta~d\phi^2)] ,  
\end{equation}
Here,~$a(t)$ represents the usual scale factor of our Universe, and since the model is flat~,~the scalar curvature is taken to be 0. Assuming that our Universe consists of matter source in the form of a barotropic perfect fluid, the modified version of the Friedmann equations for generalized Rastall gravity are given as \cite{das2018cosmological,ziaie2020structure}:
\begin{align}\label{5}
 (12\epsilon\lambda-3)H^2+6\epsilon\lambda\Dot{H} =-\epsilon\rho,\\
 \label{6}
 (12\epsilon\lambda-3)H^2+(6\epsilon\lambda -2)\Dot{H}=\epsilon p,  
\end{align}
Where $\rho$ and $p$ are the energy density and pressure~,~respectively. The Hubble parameter is given by $H\equiv\frac{\Dot{a}}{a}$, and an over-dot symbolizes a derivative with respect to the cosmic time variable $t$. Now,~from Eqn.(\ref{1})~,~the generalized version of the energy conservation equation in its explicit form is given as follows \cite{das2018cosmological}: 
\begin{equation}\label{7}
\frac{d}{dt}\left(\frac{\rho+6\lambda\Dot{H}}{1-4\epsilon\lambda}\right)+3H(\rho+p)=0.
\end{equation}
All these equations from Eqn.(\ref{5})-(\ref{7}) are dependent as we can acquire Eqn.(\ref{7}) by using Eqn.(\ref{5}) and Eqn.(\ref{6}).\\[1mm]
Converting the above modified Friedmann equations into the usual form, we get the following expressions:
\begin{equation}\label{8}
3H^2=\epsilon(\rho+\rho_{_{DE}}),    
\end{equation}
and
\begin{equation}\label{9}
2\Dot{H}=-\epsilon(\rho+\rho_{_{DE}}+p+p_{_{DE}}),   
\end{equation}
Where $p_{_{DE}}$ represents the thermodynamical pressure component originating from the generalized Rastall gravity theory and is given by the expression $p_{_{DE}}=-(12\lambda H^2+6\lambda\Dot{H})$~,~the expression of the additional energy density is also given by $\rho_{_{DE}}=12\lambda H^2+6\lambda\Dot{H}$. By the above expressions, it is clear that $p_{_{DE}}=-\rho_{_{DE}}$ which implies that the varying cosmological parameter in the Einstein gravity theory is nothing but the additional matter field component of the Rastall theory. Thus, we can conclude that Einstein's gravity theory with a varying cosmological parameter term and generalized Rastall gravity theory are identical, which further strengthens the assertion mentioned in \cite{visser2018rastall} even for this modified version of Rastall gravity theory known to be generalized Rastall gravity theory. Considering the Universe to consist of dark matter and dark energy, we can apply the conservation equation~,~Eqn.(\ref{7}) separately. Again, since dark matter is pressureless,~we get $p_{_{DM}}=0$.~So, solving Eqn.(\ref{7})~,~the expression for dark matter's energy density is given as below:
\begin{equation}\label{density_dm}
\rho_{_{DM}}=\rho_{_{DM0}}\left(1+z\right)^{3\left(\frac{R-4\epsilon\delta}{R-3\epsilon\delta}\right)},
\end{equation}
$\rho_{_{DM0}}$ is the present value of dark matter's energy density.\\
Again, using the dimensionless density parameter $\Omega_{_{m0}}=\frac{8\pi G\rho_{_{DM0}}}{3H^2_0}$~,~the above energy density equation of dark matter reduces to the following form:
\begin{equation}\label{density_DM}
\rho_{_{DM}}=\frac{3H^2_0}{8\pi G}~\Omega_{_{m0}}\left(1+z\right)^{3\left(\frac{R-4\epsilon\delta}{R-3\epsilon\delta}\right)}.    
\end{equation}\\
The expression of the corresponding Ricci scalar for generalized Rastall gravity theory is given as follows:
\begin{equation}\label{10}
R=\epsilon\left[4\delta+\left\{\rho_{_{DM}}+\left(1-3\omega(z)\right)\rho_{_{DE}}\right\}\right],    
\end{equation}
Hence,~the expression for the corresponding Hubble parameter in generalized Rastall gravity theory is given by:
\begin{widetext}
\begin{equation}\label{HPE}
 H^2=\frac{\epsilon}{3\left[\rho_{_{DM}}+\left(1-3\omega(z)\right)\rho_{_{DE}}\right]}\left[\rho_{_{DM}}^2+\left(1-3\omega(z)\right)\rho_{_{DE}}^2+\left(2-3\omega(z)\right)\rho_{_{DM}}\rho_{_{DE}}+\delta\left\{\rho_{_{DM}}-\left(2+3\omega(z)\right)\rho_{_{DE}}\right\}\right].    
\end{equation}\\
\end{widetext}
Now, let us explore our considered two recently developed dynamical dark-energy equation of state parameterization models in the context of generalized Rastall gravity theory,~such as:
\subsection*{Model-1}\label{SS1A}
 \begin{itemize}
     \item {\bf{Chaudhary-Bouali-Debnath-Roy-Mustafa-type parametrization (CBDRM)}:}
The equation of state for CBDRM parametrization is given by \cite{chaudhary2023constraints,khurana2024exploring}:
\begin{equation}\label{EOS1}
\omega(z)=\omega_0+\omega_1~\frac{(1+z)}{(2+z)},    
\end{equation}
Here, both $\omega_0$ and $\omega_1$ are parameters with constant values. Whenever the redshift increases,~that is,~$z\rightarrow \infty$,~this parametrization converges to a constant value of~$(\omega_0+\omega_1)$.~Also, for lower redshift values,~i.e.$z\rightarrow -1$,~this parametrization converges to the value~$\omega_0$. Thus, it implies that, in the past phase of our Universe, it was of a consistent nature. In addition, it shows that the state equation for dark energy was somewhat stable and unchanged and tended to the value $(\omega_0+\omega_1)$. On the other hand, in the present era, this state equation shows a tendency to that of the constant value $\omega_0$.\\
 The corresponding energy density equation is given as \cite{khurana2024exploring}:
 \begin{equation}\label{density_DE1}
  \rho_{_{DE}}=\rho_{_{DE0}}~(1+z)^{3(1+\omega_0)}(2+z)^{3\omega_{1}},   
 \end{equation}\\
Now, let us consider a dimensionless density parameter $\Omega_{_{DE0}}=\frac{8\pi G\rho_{_{DE0}}}{3H^2_0}$~,~thus the energy density equation for the model mentioned above transform into the following form:
\begin{equation}\label{density_DE1F}
\rho_{_{DE}}=\frac{3H^2_0}{8\pi G}~\Omega_{_{DE0}}~(1+z)^{3(1+\omega_0)}(2+z)^{3\omega_{1}},    
\end{equation}\\
Again, from Eqn.(\ref{HPE})~,~we get the desired form of the Hubble parameter for the CBDRM parameterization model of dark energy in the backdrop of generalized Rastall gravity theory as below:
\begin{widetext}
\begin{multline}\label{HPE1}
H^2(z)=\frac{\epsilon}{3\left[\left(1+z\right)^{3\left(\frac{R-4\epsilon\lambda}{R-3\epsilon\lambda}\right)}\Omega_{_{m0}}+\left\{1-3\left(\omega_0+\omega_1~\frac{1+z}{2+z}\right)\right\}(1+z)^{3(1+\omega_0)}(2+z)^{3\omega_{1}}\Omega_{_{DE0}}\right]} \times\\
\left[\frac{3H^2_0}{8\pi G}\left\{\left(1+z\right)^{6\left(\frac{R-4\epsilon\lambda}{R-3\epsilon\lambda}\right)}\Omega_{_{m0}}^2+\left(1-3\left(\omega_0+\omega_1~\frac{1+z}{2+z}\right)\right)(1+z)^{6(1+\omega_0)}(2+z)^{6\omega_{1}}\Omega_{_{DE0}}^2+\left(2-3\left(\omega_0+\omega_1~\frac{1+z}{2+z}\right)\right)\right.\ \right.\\
\left.\ \left.\left(1+z\right)^{3\left(\frac{R-4\epsilon\lambda}{R-3\epsilon\lambda}+1+\omega_0\right)}(2+z)^{3\omega_{1}}\Omega_{_{m0}}\Omega_{_{DE0}}\right\}+\delta\left\{\left(1+z\right)^{3\left(\frac{R-4\epsilon\lambda}{R-3\epsilon\lambda}\right)}\Omega_{_{m0}}-\left(2+3\left(\omega_0+\omega_1~\frac{1+z}{2+z}\right)\right)\right.\right. \\
\left.\left.(1+z)^{3(1+\omega_0)}(2+z)^{3\omega_{1}}\Omega_{_{DE0}}\right\}\right].          
\end{multline}    
\end{widetext}
\end{itemize}
\subsection*{Model-2}\label{SS1B}
 \begin{itemize}
  \item{\bf{Chaudhary-Debnath-Mustafa-Maurya-Atamurotov-type parametrization (CDMMA)}:}
The equation of state for CDMMA parametrization is given by \cite{chaudhary2024addressing,khurana2024exploring}:
\begin{equation}\label{EOS2}
 \omega(z)=\omega_{0}+\frac{\alpha+(1+z)^{\beta}}{\omega_{1}+\omega_{2}(1+z)^{\beta}},    
\end{equation}\\
All of the terms~$\omega_{0}$~,~$\omega_{1}$~,~$\omega_{2}$~,~$\alpha$~,~$\beta$~ present in the parametrization as mentioned above are constants that strongly indicate towards both the convergence and divergence nature of it. Whenever these constants obey certain kinds of conditions, this parameterization shows a consistent behavior in the ancient past phase of our Universe, that is, as $z\rightarrow \infty$,~it gives the value $(\omega_{0}+\frac{1}{\omega_{2}})$. But in the case of the present epoch,~i.e.,~for $z\rightarrow -1$,~this parameterization shows either a convergent nature tending to the value~$(\omega_{0}+\frac{\alpha}{\omega_{1}})$~when the constant~$\beta>0$~or a divergent nature whenever the limit cannot be defined or can approach a completely different value depending upon the values of these parameters. This indicates that the parameterization mentioned above is sensitive to the choice of the parameters. Thus, we can quickly conclude that in this particular parameterization, the selection of the parameters is crucial for the accuracy of the equation of the state of the dark energy across different cosmic epochs.  
The corresponding energy density equation is given as \cite{khurana2024exploring,chaudhary2024addressing}:
\begin{equation}\label{density_DE2}
\rho_{_{DE}}=\rho_{_{DE0}}~(1+z)^{3\left(1+\omega_{0}+\frac{\alpha}{\omega_{1}}\right)}\left[\omega_{1}+\omega_{2}(1+z)^{\beta}\right]^{\frac{3(\omega_{1}-\alpha \omega_{2})}{\beta \omega_{1}\omega_{2}}}, 
\end{equation}\\
In terms of some dimensionless density parameter $\Omega_{_{DE0}}=\frac{8\pi G\rho_{_{DE0}}}{3H^2_0}$~,~it reduces to the following form as given below:
\begin{equation}\label{density_DE2F}
\rho_{_{DE}}=\frac{3H^2_0}{8\pi G}~\Omega_{_{DE0}}~(1+z)^{3\left(1+\omega_{0}+\frac{\alpha}{\omega_{1}}\right)}\left[\omega_{1}+\omega_{2}(1+z)^{\beta}\right]^{\frac{3(\omega_{1}-\alpha \omega_{2})}{\beta \omega_{1}\omega_{2}}},    
\end{equation}\\
Finally, using the Eqn.(\ref{HPE})~,~we get our desired expression of the Hubble parameter for the CDMMA parametrization model of dark energy in the backdrop of generalized Rastall gravity theory as follows:
\begin{widetext}
\begin{multline}\label{HPE2}
H^2(z)=\frac{\epsilon}{3\left[\left(1+z\right)^{3\left(\frac{R-4\epsilon\lambda}{R-3\epsilon\lambda}\right)}\Omega_{_{m0}}+\left\{1-3\left(\omega_{0}+\frac{\alpha+(1+z)^{\beta}}{\omega_{1}+\omega_{2}(1+z)^{\beta}}\right)\right\}(1+z)^{3\left(1+\omega_{0}+\frac{\alpha}{\omega_{1}}\right)}[\omega_{1}+\omega_{2}(1+z)^{\beta}]^{\frac{3(\omega_{1}-\alpha \omega_{2})}{\beta \omega_{1}\omega_{2}}}\Omega_{_{DE0}}\right]} \times\\
\left[\frac{3H^2_0}{8\pi G}\left\{\left(1+z\right)^{6\left(\frac{R-4\epsilon\lambda}{R-3\epsilon\lambda}\right)}\Omega_{_{m0}}^2+\left(1-3\left(\omega_{0}+\frac{\alpha+(1+z)^{\beta}}{\omega_{1}+\omega_{2}(1+z)^{\beta}}\right)\right)(1+z)^{6\left(1+\omega_{0}+\frac{\alpha}{\omega_{1}}\right)}\left[\omega_{1}+\omega_{2}(1+z)^{\beta}\right]^{\frac{6(\omega_{1}-\alpha \omega_{2})}{\beta \omega_{1}\omega_{2}}}\Omega_{_{DE0}}^2\right.\ \right.\\
\left.\ \left.+\left(2-3\left(\omega_{0}+\frac{\alpha+(1+z)^{\beta}}{\omega_{1}+\omega_{2}(1+z)^{\beta}}\right)\right)
\left(1+z\right)^{3\left(\frac{R-4\epsilon\lambda}{R-3\epsilon\lambda}+1+\omega_{0}+\frac{\alpha}{\omega_{1}}\right)}\left[\omega_{1}+\omega_{2}(1+z)^{\beta}\right]^{\frac{3(\omega_{1}-\alpha \omega_{2})}{\beta \omega_{1}\omega_{2}}}\Omega_{_{m0}}\Omega_{_{DE0}}\right\}\right.\\
\left.+\delta\left\{\left(1+z\right)^{3\left(\frac{R-4\epsilon\lambda}{R-3\epsilon\lambda}\right)}\Omega_{_{m0}}-\left(2+3\left(\omega_{0}+\frac{\alpha+(1+z)^{\beta}}{\omega_{1}+\omega_{2}(1+z)^{\beta}}\right)\right)(1+z)^{3\left(1+\omega_{0}+\frac{\alpha}{\omega_{1}}\right)}\left[\omega_{1}+\omega_{2}(1+z)^{\beta}\right]^{\frac{3(\omega_{1}-\alpha \omega_{2})}{\beta \omega_{1}\omega_{2}}}\Omega_{_{DE0}}\right\}\right]. 
\end{multline}
\end{widetext}
\end{itemize}


\section{Methodology and Data Description}\label{S2}
\begin{figure*}
\centering
\includegraphics[width=18.0 cm]{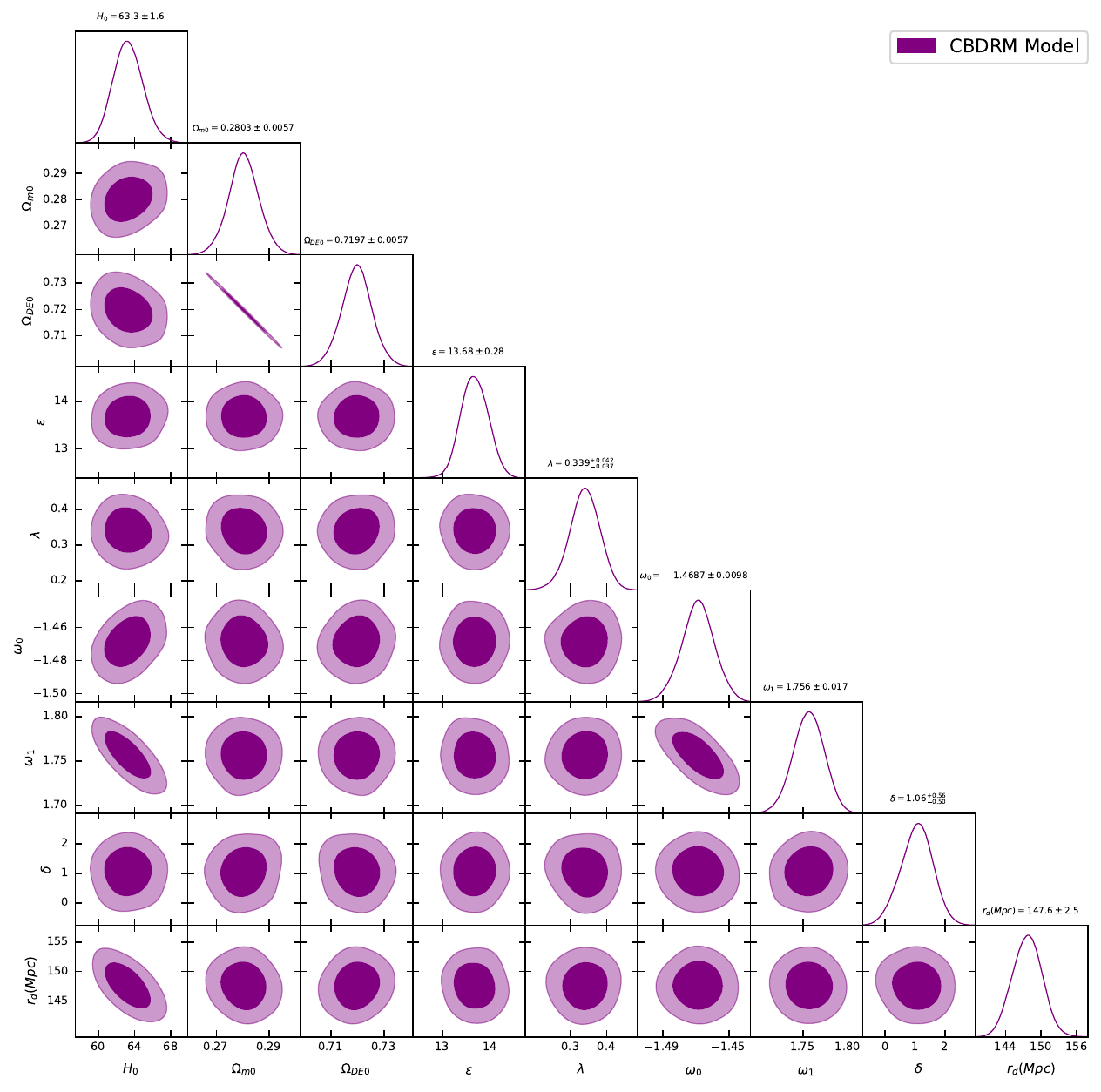}
\caption{The posterior distributions of the CBDRM model at 68\% (1\(\sigma\)) and 95\% (2\(\sigma\)) credible intervals}\label{fig_1}
\end{figure*}

\begin{figure*}
\centering
\includegraphics[width=18.0 cm]{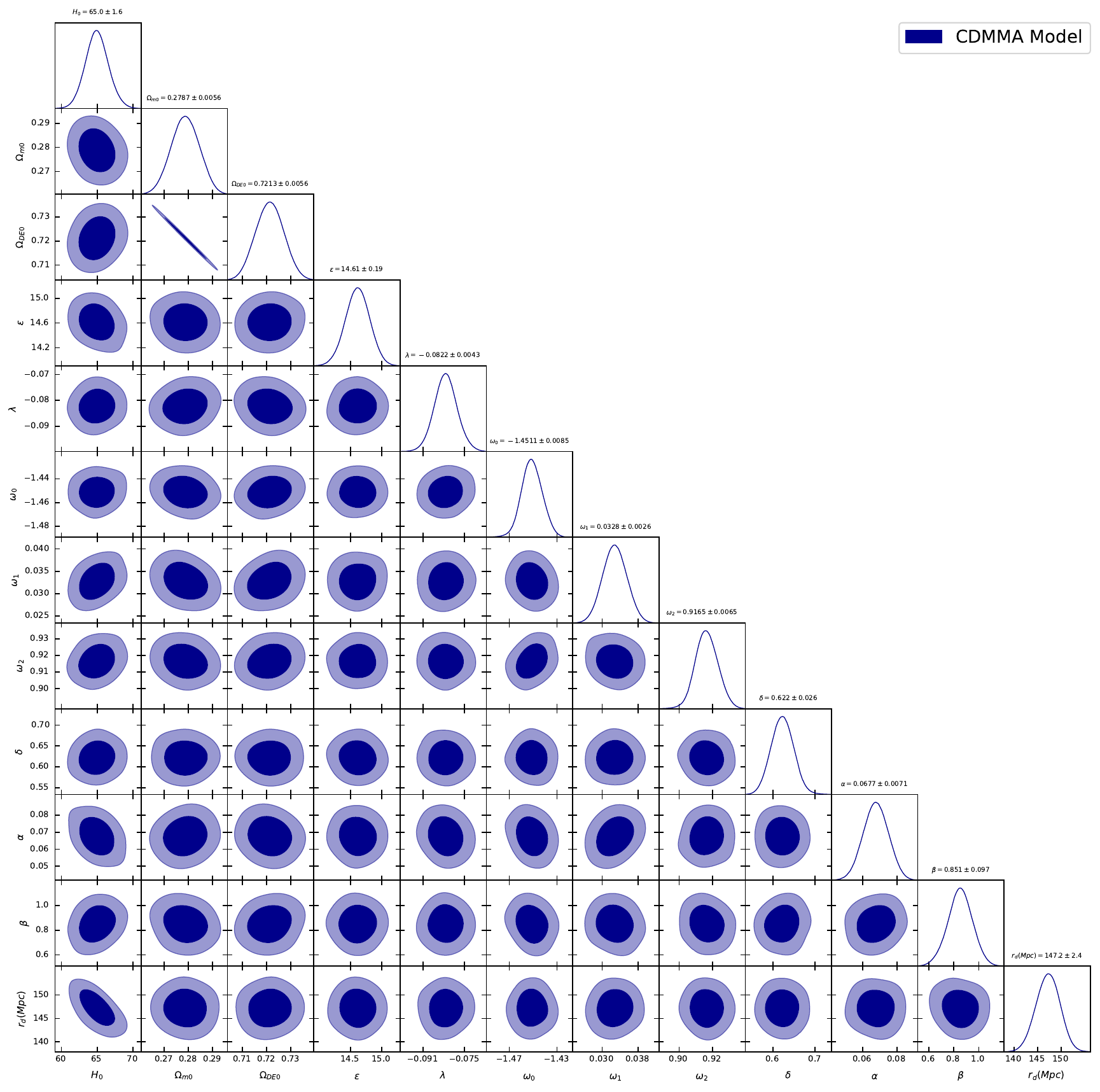}
\caption{The posterior distributions of the CDMMA model at 68\% (1\(\sigma\)) and 95\% (2\(\sigma\)) credible intervals}\label{fig_2}
\end{figure*}

In this study, we use the \texttt{polyChord} nested sampling algorithm to estimate the distribution of parameter values for the CBDRM and CDMMA models, analyzing changes in black hole mass during the evolution of the Universe. \texttt{polyChord} is a Bayesian technique that iteratively refines a set of live points to explore the parameter space and compute posterior distributions. The posterior is expressed as:
\begin{equation}
P(\theta|D) = \frac{L(D|\theta) P(\theta)}{P(D)},
\end{equation}
where \(P(\theta|D)\) is the posterior probability, \(L(D|\theta)\) is the likelihood, \(P(\theta)\) is the prior, and \(P(D)\) is the evidence. Using data sets such as cosmic chromatometer observations and Type Ia supernovae data (without SHOES calibration), we defined a model with parameters \(\theta\), specified informed priors, and initialized the live points. The sampling process iteratively refined the posterior distribution, allowing the distribution of parameters and offering a comprehensive approach to studying black hole mass evolution within cosmological models.
\begin{itemize}
    \item Cosmic Chronometers: In this study, we use 31 Cosmic Chronometers (CC) measurements spanning a redshift range \(0 \leq z \leq 1.965\) \cite{jimenez2003constraints,simon2005constraints,simon2010,moresco2012new,zhang2014four,moresco20166,ratsimbazafy2017age}, obtained via the differential age method \cite{jimenez2002constraining}. This technique calculates the Hubble parameter by measuring the rate of change of redshift with respect to cosmic time (\(\Delta z / \Delta t\)) in passively evolving galaxies, providing a model-independent estimate of the Universe's expansion rate. The data points are sourced from various independent studies, as detailed in Table 1 of \cite{vagnozzi2021eppur}. For our MCMC analysis, we assess the goodness-of-fit using the \(\chi^2_{CC}\) statistic: $\chi^2_{CC}(\theta) = \Delta H^T(z) \mathbf{C}^{-1} \Delta H(z),$ where \( \Delta H(z) = \mathbf{H_{\text{model}}}(\theta) - \mathbf{H_{\text{obs}}} \) is the residual vector of the model and observed Hubble parameters. \( \mathbf{H_{\text{model}}}(\theta) \) represents the theoretical Hubble parameter in redshift \( z_i \) for the model parameters \(\theta\), and \( \mathbf{H_{\text{obs}}} \) is the vector of observed values. The covariance matrix \( \mathbf{C} \), with diagonal elements \( \sigma_H^2(z_i) \), accounts for observational uncertainties, assuming uncorrelated data points. The inverse of this matrix, \( \mathbf{C}^{-1} \), is used in the fit to incorporate measurement errors.
    \item The Pantheon$^{+}$ compilation contains light curves for 1701 Type Ia supernovae (SNe Ia), extracted from 1550 different events, and spans a redshift range of \(0 \leq z \leq 2.3\) \cite{brout2022pantheon}. The key observable for SNe Ia is the apparent magnitude, expressed as $m(z) = 5 \log_{10} \left( \frac{D_L (z)}{\text{Mpc}} \right)  + 25 + \mathcal{M},$ where \(\mathcal{M}\) denotes the absolute magnitude of the SNe Ia. The luminosity distance \(D_L\) in a flat FLRW cosmology is given by \cite{brout2022pantheon}: $D_L(z) = c(1 + z) \int_{0}^{z} \frac{\mathrm{d}z'}{H(z')},$ with \(c\) denoting the speed of light. To evaluate the agreement between observed data and theoretical model predictions, the \(\chi^2\) statistic for SNe Ia is defined as $\chi^2_{\text{SNe Ia}} = \Delta \mathbf{d}^T \mathbf{C}_{\text{tot}}^{-1} \Delta \mathbf{d},$ where \(\mathbf{C}_{\text{tot}} = \mathbf{C}_{\text{stat}} + \mathbf{C}_{\text{sys}}\) combines the statistical (\(\mathbf{C}_{\text{stat}}\)) and systematic (\(\mathbf{C}_{\text{sys}}\)) covariance matrices, and \(\Delta \mathbf{d}\) is the vector of differences between the observed and model distance moduli. In our analysis, marginalization over \(\mathcal{M}\) is performed \cite{escobal2021cosmological}, equivalent to minimizing \(\chi^2\) with respect to \(\mathcal{M}\). This procedure leads to the projected chi-square: $\chi^2_{\text{SNe Ia,Proj}} = A(z_{\text{i}}, \theta) - \frac{B(z_{\text{i}}, \theta)^2}{C(z_{\text{i}}, \theta)},$ where $A(z_{\text{i}}, \theta) = \sum_{i=1}^{N} \frac{(\mu_{z_i} - \mu(z_{\text{i}}, \theta))^2}{\sigma_{\text{Data}}^2(z_i)}$, $B(z_{\text{i}}, \theta) = \sum_{i=1}^{N} \frac{\mu_{z_i} - \mu(z_{\text{i}}, \theta)}{\sigma_{\text{Data}}^2(z_i)}$, and 
    $C(z_{\text{i}}, \theta) = \sum_{i=1}^{N} \frac{1}{\sigma_{\text{Data}}^2(z_i)}.$    
    \item We also incorporate the latest Baryon Acoustic Oscillation (BAO) measurements from the Dark Energy Spectroscopic Instrument (DESI) Year 1 \cite{adame2024desi} and the Dark Energy Survey (DES) Year 6 \cite{abbott2024dark}. BAO measurements depend on the sound horizon at baryon decoupling (\(r_d\)), around \(z_d \approx 1060\) in the standard model. It is given by $r_d = \int_{z_d}^{\infty} \frac{c_s(z)}{H(z)} \, dz,$ Here, \(c_s(z)\) represents the sound speed, and \(H(z)\) is the Hubble parameter. In the flat \(\Lambda\)CDM model, the sound horizon is \(r_d = 147.09 \pm 0.26 \, \text{Mpc}\) \cite{collaboration2020planck}. To adopt a model-independent approach, we consider \(r_d\) as a free parameter, avoiding assumptions about early Universe physics or specific recombination models \cite{pogosian2020recombination,jedamzik2021reducing,pogosian2024consistency,lin2021early,vagnozzi2023seven}. This strategy enables late-time cosmological data to directly constrain \(r_d\) alongside other parameters. To extract posterior distributions of each model using the BAO measurements, we compute the Hubble distance \(D_H(z)\), the comoving angular diameter distance \(D_M(z)\), and the volume-averaged distance \(D_V(z)\), defined as: $D_H(z) = \frac{c}{H(z)}, D_M(z) = c \int_0^z \frac{dz'}{H(z')},\text{and} D_V(z) = \left[ z D_M^2(z) D_H(z) \right]^{1/3}.$ For the analysis, we focus on the ratios \( \frac{D_M(z)}{r_d} \), \( \frac{D_H(z)}{r_d} \), and \( \frac{D_V(z)}{r_d} \). The \(\chi^2\) statistic for these ratios is expressed as: $\chi^2_{D_Y / r_d} = \Delta D_Y^T \cdot \mathbf{C}^{-1}_{D_Y} \cdot \Delta D_Y,$ where \( \Delta D_Y = D_{Y / r_d, \text{Model}} - D_{Y / r_d, \text{Data}} \) for \( Y = H, M, V \), and \( \mathbf{C}^{-1}_{D_Y} \) represents the inverse of the covariance matrix. The covariance matrix \( \mathbf{C}_{D_Y} \) is generally diagonal with elements \( \sigma_{D_Y}^2 \), representing the uncertainties, i.e., \( \mathbf{C}_{D_Y} = \text{diag}(\sigma_{D_Y}^2) \), and its inverse is \( \mathbf{C}^{-1}_{D_Y} = (\text{diag}(\sigma_{D_Y}^2))^{-1} \). The total \(\chi^2\) for the BAO dataset is then: $\chi^2_{\text{BAO}} = \chi^2_{D_H / r_d} + \chi^2_{D_M / r_d} + \chi^2_{D_V / r_d}.$ This total \(\chi^2_{\text{BAO}}\) accounts for the contributions from the Hubble distance, angular diameter distance, and volume-averaged distance.
\end{itemize}
To combine information from multiple datasets, the overall chi-squared statistic is defined as the sum of the contributions from all datasets:  
\[
\chi^2_{\text{tot}} = \chi^2_{\text{CC}} + \chi^2_{\text{SNe Ia,Proj}} + \chi^2_{\text{BAO}}.
\]
In our analysis, parameter estimation and Bayesian inference are performed using the Markov Chain Monte Carlo (MCMC) framework provided by the \texttt{polyChord} library \cite{handley2015polychord}. For visualizing and summarizing posterior distributions, the \texttt{GetDist} package \cite{lewis2019getdist} is utilized, offering tools for generating both 1D and 2D posterior density plots. Figs. ~\ref{fig_1} and \ref{fig_2} show the triangle plots of the CBDRM and CDMMA models, displaying the marginal distributions of each parameter along the diagonal, while the off-diagonal elements represent the 2D joint distributions between pairs of parameters. Table ~\ref{tab_1} presents the mean values of each parameter along with their corresponding 68\% (1\(\sigma\)) and 95\% (2\(\sigma\)) credible intervals and prior ranges. In our analysis, the $\Lambda$CDM model predicts a present-day Hubble constant of \(H_0 = 67.7 \pm 1.1\), which is consistent with the Planck Collaboration's estimate and aligns with the DESI Collaboration's prediction. The matter-energy densities in the $\Lambda$CDM model (\(\Omega_{m0} = 0.315 \pm 0.007\) and \(\Omega_{\Lambda 0} = 0.685 \pm 0.007\)) closely match the Planck results, and the predicted value of the sound horizon, \(r_d\), is also in good agreement with Planck's estimate. For the CBDRM model, the predicted Hubble constant is \(H_0 = 63.3 \pm 1.6\), while for the CDMMA model, it is \(H_0 = 65.0 \pm 1.6\), both deviating from the Planck predictions. However, the predicted values of \(\Omega_{m0}\) are close to those of the DESI Collaboration, and the estimates for \(r_d\) in both cases remain consistent with Planck's results. Further to determine which model is more favorable, we compare CBDRM and CDMMA against the \(\Lambda\)CDM model using statistical tools that evaluate fit and complexity. The minimum chi-squared value, \(\chi^2_{\text{min}}\), is obtained via maximum likelihood estimation, with the likelihood function \(\mathcal{L}_{\text{tot}} = e^{-\chi^2_{\text{tot}}/2}\). The reduced chi-squared statistic, \(\chi_{\text{red}}^2 = \chi_{\text{min}}^2/\text{DOF}\), where DOF is the difference between data points and model parameters, indicates a good fit when close to 1 \cite{andrae2010and}. Model comparison uses the Akaike Information Criterion (AIC) and Bayesian Information Criterion (BIC) \cite{kuha2004aic,liddle2007information,akaike1974new,burnham2004multimodel,schwarz1978estimating}, defined as: $\text{AIC} = \chi^2_{\text{min}} + 2k_{\text{tot}}, \quad \text{BIC} = \chi^2_{\text{min}} + k_{\text{tot}}\ln(\mathcal{N}_{\text{tot}})$
where \(k_{\text{tot}}\) is the number of parameters and \(\mathcal{N}_{\text{tot}}\) is the dataset size. Differences with respect to \(\Lambda\)CDM are: $\Delta \text{AIC} = \text{AIC}_{\text{CBDRM/CDMMA}} - \text{AIC}_{\Lambda \text{CDM}}, \quad \Delta \text{BIC} = \text{BIC}_{\text{CBDRM/CDMMA}} - \text{BIC}_{\Lambda \text{CDM}}$ According to Jeffreys’ scales \cite{jeffreys1998theory}, \(|\Delta \text{AIC}|\leq 2\) indicates comparable models, while \(\geq 4\) favors the model with the lower AIC. For BIC, differences between 2 and 6 suggest strong evidence against the model, and $>$ 6 indicates very strong evidence. Negative \(\Delta \text{AIC}\) or \(\Delta \text{BIC}\) favors CBDRM or CDMMA. Statistical significance is assessed via the p-value: $p = 1 - F_{\chi^2_{\text{min}}}(\chi \mid \nu)$
where \(F_{\chi^2}\) is the chi-squared CDF and \(\nu\) denotes degrees of freedom. A p-value \(< 0.05\) indicates strong evidence against the null hypothesis \cite{wasserstein2016asa}. Table \ref{tab_2} shows that the CDMMA model is the most favorable according to the AIC criterion, which is less strict regarding model complexity. However, if simplicity is prioritized, as reflected in the BIC values, the \(\Lambda\)CDM model is preferred. Overall, the CDMMA model provides the best balance of fit quality and statistical significance, making it the most suitable choice when the increased complexity is acceptable.
\begin{table}
\begin{tabular}{|c|c|c|c|}
\hline
Models & Parameter & Prior & JOINT  \\
\hline
& $H_{0}$ &$[50.,90.]$   &$67.733^{\pm 1.157}_{\pm 2.226}$ \\
$\Lambda$CDM Model& $\Omega_{mo}$ &$[0,1]$   &$0.328^{\pm 0.009}_{\pm 0.017}$ \\
& $\Omega_{\Lambda0}$ &$[0,1]$   &$0.671^{\pm 0.008}_{\pm 0.018}$ \\
& $r_d$ (Mpc) & $[100,200]$ &$147.150^{\pm 2.517}_{\pm 4.768}$ \\
\hline
& $H_{0}$ &$[50.,90.]$   &$63.282{^{\pm 1.623}_{\pm 2.982}}$ \\
& $\Omega_{mo}$ &$[0,1]$   &$0.280{^{\pm 0.005}_{\pm 0.011}}$ \\
& $\Omega_{DE0}$ &$[0,1]$   &$0.719{^{\pm 0.005}_{\pm 0.011}}$ \\
& $\epsilon$ &$[13.,15.]$   &$13.678{^{\pm 0.273}_{\pm 0.542}}$ \\
CBDRM Model & $\lambda$ &$[0.,0.1]$   &$0.339{^{\pm 0.037}_{\pm 0.086}}$ \\
&$\omega_{0}$   &$[-2.,-1.]$  &$-1.469{^{\pm 0.099}_{\pm 0.020}}$  \\
&$\omega_{1}$   &$[0.,0.1]$  &$1.756{^{\pm 0.017}_{\pm 0.036}}$  \\
&$\delta$   &$[0.,1.]$  &$1.059{^{\pm 0.540}_{\pm 1.058}}$  \\
&$r_d$ (Mpc) & $[100,200]$ &$147.637{^{\pm 2.630}_{\pm 4.610}}$ \\
\hline 
& $H_{0}$ &$[50.,90.]$   &$65.032{^{\pm 1.460}_{\pm 2.969}}$ \\
& $\Omega_{mo}$ &$[0,1]$   &$0.278{^{\pm 0.006}_{\pm 0.011}}$ \\
& $\Omega_{DE0}$ &$[0,1]$   &$0.721{^{\pm 0.006}_{\pm 0.011}}$ \\
& $\epsilon$ &$[13.,15.]$   &$14.614{^{\pm 0.175}_{\pm 0.394}}$ \\
& $\lambda$ &$[0.,0.1]$   &$-0.082{^{\pm 0.004}_{\pm 0.008}}$ \\
CDMMA Model &$\omega_{0}$   &$[-2.,-1.]$  &$-1.451{^{\pm 0.008}_{\pm 0.014}}$  \\
&$\omega_{1}$   &$[0.,0.1]$  &$0.033{^{\pm 0.003}_{\pm 0.005}}$  \\
&$\omega_{2}$   &$[0.,1.]$  &$0.916{^{\pm 0.006}_{\pm 0.012}}$  \\
&$\delta$   &$[0.,1.]$  &$0.622{^{\pm 0.026}_{\pm 0.051}}$  \\
&$\alpha$   &$[0.,0.1]$  &$0.068{^{\pm 0.007}_{\pm 0.013}}$  \\
&$\beta$   &$[0.,1.]$  &$0.851{^{\pm 0.093}_{\pm 0.201}}$  \\
&$r_d$ (Mpc) & $[100,200]$ &$147.185{^{\pm 2.318}_{\pm 4.798}}$ \\
\hline
\end{tabular}
\caption{Mean values, along with 68\% (1\(\sigma\)) and 95\% (2\(\sigma\)) credible intervals, and prior ranges for parameters of the $\Lambda$CDM, CBDRM, and CDMMA models.}\label{tab_1}
\end{table}
\begin{table*}[htbp]
\begin{center}
\begin{tabular}{|c|c|c|c|c|c|c|c|c|c|}
\hline
Models & ${\chi_{\text{tot},min}^2}$ & $N_{tot}$ & $k$ & $\chi_{\text {red}}^2$ & AIC & $\Delta$AIC & BIC & $\Delta$BIC & p-value \\[0.1cm]
\hline
$\Lambda$CDM Model & 1731.12 & 1745 & 4 & 0.994 & 1739.12 & 0 & 1760.97 & 0 & 0.562 \\[0.1cm] 
\hline
CBDRM Model & 1712.72 & 1745 & 9 & 0.986 &1730.72  & -8.39 & 1779.90 & 18.92  & 0.650 \\
\hline
CDMMA Model & 1705.11 & 1745 & 12 & 0.983 & 1729.11 & -10.00 & 1794.68 & 33.70 & 0.679  \\
\hline
\end{tabular}
\caption{Statistical Metrics for $\Lambda$CDM, CBDRM, and CDMMA Models including \(\chi_{\text{tot,min}}^2\), \(\chi^2_{\text{red}}\), AIC, BIC, \(\Delta\mathrm{AIC}\), \(\Delta\mathrm{BIC}\) and p-values}\label{tab_2}
\end{center}
\end{table*}


\section{Non-singular Black holes and their Mass Accretion Process in the surroundings of Generalized Rastall Gravity}\label{S3}
The mass accretion process of a black hole is very significant from the perspective of not only cosmological studies but is quite important in astrophysics and fundamental physics. In this paper, we choose to study the mass accretion of non-singular black holes in the context of generalized Rastall gravity, which seems to be a great way to avoid singularities with the help of modified gravity theory, leading to a unique way to study the dynamics of accretion. This process can further advance research into the thermodynamic properties of black holes and enables researchers to calculate their radiation emission rates. Here, we study this process with the help of the constrained values of the parameters present in Table \ref{tab_1} obtained in the previous Section \ref{S2}~through the MCMC technique to observe the effects of these constrained values on the mass accretion of the black holes. Also, this mass-accretion process can give us insight into how black holes will behave during the evolution of the Universe. These results can be tested against observational data, providing a powerful tool to probe the structure of black holes and to analyze the nature of gravity.\\\\
In this particular section, let us first discuss the non-singular black hole solution and its properties in detail against the backdrop of the generalized Rastall gravity theory. Then, we are going to develop its mass accretion process.
\subsection*{Non-singular Black Holes and their properties}\label{SS3A}
In the context of generalized Rastall gravity, let us consider an approach of changing the Schwarzschild singularity into a de Sitter vacuum following the idea of Sakharov \cite{sakharov1966initial} that in the equation of state of super-high density, we can take a negative density. According to the perspective of Gliner \cite{gliner1966algebraic}~,~this negative density corresponds to a vacuum~,~and it finally reduces to the state of gravitational collapse. This kind of approach was reflected in several works such as \cite{dymnikova1992vacuum,dymnikova2002cosmological,dymnikova2003spherically,moradpour2019black}. So, according to \cite{dymnikova1992vacuum,moradpour2019black} the energy density is given by:
\begin{equation}\label{bh.1}
 \sigma(r)=\mathfrak{a}~exp\left(-\frac{r^3}{\mathfrak{b}^3}\right), 
\end{equation}
Generally, $\mathfrak{a}$~and~$\mathfrak{b}$ are taken as unknown constants. Now, let us assume a general form of a static and spherically symmetric black hole metric given as \cite{dutta2019dark,debnath2015accretions,debnath2015accretion}:
\begin{equation}\label{bh.2}
 ds^2=-f(r)dt^2+\frac{dr^2}{f(r)}+r^2(d\theta^2+sin^2\theta d\phi^2),   
\end{equation}
Again, the field equation given by Eqn.(\ref{3}) reduces to the following form \cite{moradpour2019black}:
\begin{equation}\label{bh.3}
G_{\mu\nu}=\epsilon T_{\mu\nu}-\epsilon\delta g_{\mu\nu}=\epsilon(T_{\mu\nu}-\delta g_{\mu\nu}),    
\end{equation}
The above equation validates the point that the term given as~$\epsilon\delta$~represents a cosmological constant, implying the fact that the anti-de Sitter spacetime is nothing but a vacuum solution \cite{moradpour2017generalization}.
The solution of Eqn.(\ref{bh.3}) leads us to \cite{moradpour2019black}:
\begin{equation}\label{bh.4}
f(r)=1-\frac{\mathcal{C}_1}{r}-\frac{\epsilon\delta}{3}r^2+\frac{\epsilon\mathfrak{a}\mathfrak{b}^3}{3r}~exp\left(-\frac{r^3}{\mathfrak{b}^3}\right),   
\end{equation}
$\mathcal{C}_1$~being the the mass parameter. Its corresponding radial and transverse pressure components are given by: 
\begin{equation}\label{metric_density}
p_{_{rad}}(r)=-\sigma(r)~and~p_{_{trans}}(r)=\left(\frac{3r^3}{2\mathfrak{b}^3}-\mathfrak{a}\right)\sigma(r),    
\end{equation}
Which clearly shows that this is an isotropic source. Again, the Ricci scalar and Ricci square associated with the above solution are in accordance with the limit $r\to 0$. If we take some particular values for $\mathcal{C}_1$~and~$\epsilon$ as follows: 
\begin{equation}\label{bh.5}
\mathcal{C}_1=\frac{\mathfrak{b}^3}{3\epsilon\mathfrak{a}}~and~\epsilon=\pm\frac{1}{\mathfrak{a}}, 
\end{equation}
Then, the Riemann and Weyl squares also comply with the above solution. Furthermore, by the positive value of $\mathcal{C}_1$ and the weak energy condition, we can ensure positivity for both the unknown constant $\mathfrak{a}$~and~$\mathfrak{b}$. As entropy is always positive, by the relation $S=\frac{2\pi}{\epsilon}A$~given in \cite{moradpour2020generalized}~,~here we can easily exclude the value $\epsilon=-\frac{1}{\mathfrak{a}}$. Again, to show that the black hole metric solution given in Eqn.(\ref{bh.4}) is everywhere non-singular together with the constraints, Eqn.(\ref{bh.5}), we can make use of the Eddington-Finkelstein coordinate transformation \cite{moradpour2019black}. This black hole solution is quite different in asymptotic nature from \cite{dymnikova1992vacuum,dymnikova2002cosmological,dymnikova2003spherically} whenever $r$ takes large values, making it competent with the Schwarzschild metric. However, smaller values of $r$ coincide with the de Sitter metric. As we can contemplate the same vacuum stress-energy momentum tensor given by Eqn.(\ref{bh.1}) and Eqn.(\ref{metric_density}) as in \cite{dymnikova1992vacuum,dymnikova2002cosmological,dymnikova2003spherically}, the internal nature for both the cases are the same, although the metric solution given by Eqn.(\ref{bh.4}) shows a de Sitter asymptote with a cosmological term $\frac{\epsilon\delta}{3}$ which is a result of non-minimal coupling property related to the background theory instead of a Schwarzschild asymptote. The metric function Eqn.(\ref{bh.4}) has a solution at the point $r=\mathfrak{b}$~,~if \cite{moradpour2019black} 
\begin{equation}\label{bh.6}
\delta=\mathfrak{a}\left(\frac{3-\mathfrak{b}^2(1-e^{-1})}{\mathfrak{b}^2}\right).    
\end{equation}
The above equation shows the relationship between the energy source, i.e., the black hole properties and the non-minimal coupling parameter $\delta$, or the varying Rastall parameter $\lambda$. Also, on the black hole horizon, that is, at $r=\mathfrak{b}$, the black hole metric $f(r)$ changes its signature from $(-,+,+,+)$ to $(+,-,+,+)$ as the metric takes a positive value for $f(r<\mathfrak{b})$ and a negative value for $f(r>\mathfrak{b})$ on both sides of the black hole horizon ($r=\mathfrak{b}$).


\subsection*{Mass accretion in generalized Rastall gravity theory}\label{SS3B}
In the process of analyzing the phenomena of mass accretion of a 4-dimensional non-singular, static, and spherically symmetric black hole metric given by Eqn.(\ref{bh.2}) together with the function $f(r)$ given by Eqn.(\ref{bh.4}), a function of $r$, let us consider $M$ to be the black hole mass. The accreting fluid or we can say the dark energy candidate, is of the form of a perfect fluid with the pressure $p$ and energy density $\rho$,~so its energy-momentum tensor is of the following form:
\begin{equation}\label{ac.1}
 T_{_{\zeta\eta}}=(\rho+p)u_{_{\zeta}} u_{_{\eta}}+pg_{_{\zeta\eta}}~~, 
\end{equation}
Where, $u^{\zeta}$ is the fluid flow four-velocity vector given as $u^{\zeta}=\frac{dx^{\zeta}}{ds}=(u^0,u^1,0,0)$~,~here only the components $u^0$ and $u^1$ are non-zero as per the spherical symmetry property of the black hole metric satisfying a certain normalize condition $u^{\zeta}u_{_{\zeta}}=-1$~,~which implies $u^0=\frac{\sqrt{f(r)+(u^1)^2}}{f(r)}$. Taking the radial velocity of the flow as $u^1=u$~,~we get $u_0=g_{_{00}}u^0=\sqrt{f(r)+u^2}$~,~where~$\sqrt{-g}=r^2sin\theta$. From Eqn.(\ref{ac.1})~,~it follows that $T^1_0=(\rho+p)uu_0$. Also, let us take the fluid flow towards the black hole, that is~,~$u<0$.\\[1.5mm]
Again, in the backdrop of generalized Rastall gravity theory, the energy-momentum conservation law is violated. So, for this particular gravity theory, the modified version of this law is given as $T^{_{\zeta\eta}}_{_{;\eta}}=\lambda R^{_{,\zeta}}$. Now, its redial temporal component part can be written as $ \frac{d}{dr}\left(T^1_0\sqrt{-g}\right)=\lambda R^{_{,1}}\sqrt{-g}$~,~called the first integral of motion~,~which turns into the following form:
\begin{equation}\label{ac.2}
\left(\rho+p\right)ur^2M^{-2}\sqrt{f(r)+u^2}=\lambda\mathcal{B}_0 +\mathcal{B}_1~,      
\end{equation}
where $\mathcal{B}_0=\left(\frac{4f(r)}{r}-2rf''(r)\right)$~and~$\mathcal{B}_1$ being any integration constant whose dimension is the same as energy density. In addition,~for this particular gravity theory,~the modified version of the energy-flux equation is of the form $u_{_{\zeta}}T^{_{\zeta\eta}}_{_{;\eta}}=\lambda u_{_{\zeta}}R^{_{,\zeta}}$~implies~$u^{_{\zeta}}\rho_{_{,\eta}}+\left(\rho+p\right)u^{_{\zeta}}_{_{;\zeta}}=\lambda u^{_{\zeta}}R_{_{,\eta}}$~.~So when~$\zeta=1$~,~we get the energy flux equation or the second integral of motion as follows:
\begin{equation}\label{ac.3}
ur^2M^{-2}exp\left[\int_{\rho_\infty}^{\rho}{\frac{d\rho}{\rho+p(\rho)}}\right]=\lambda ur^2\mathcal{B}_0-\mathcal{B}_2~,    
\end{equation}
With the positive energy flux-related integration constant $\mathcal{B}_2$, and the minus sign is taken as per our convenience. Also, $\rho_\infty$ and $\rho$ are the energy densities of the black hole at infinity and at the black hole horizon, respectively. 
We can compute the rate of mass change for the black hole by integrating the dark energy flux into the 4-dimensional volume of the black hole,~ using the formula,~$\Dot{M}=-\int T^1_0 dS$~where~$dS=\sqrt{-g}d\theta d\phi$.
So, the final form of the required rate of change of mass function is given by:
\begin{equation}\label{mass_rate}
\Dot{M}=4\pi\left(\lambda\mathcal{B}_0+\mathcal{B}_1\right)\left(\rho+p(\rho)\right)M^2,     
\end{equation}
The above equation shows the dependency of the changes in the mass of the black hole on the particular term $\rho+p$. Thus, in the phantom-filled Universe~($\rho+p<0$),~the mass of the black hole will gradually decrease; on the contrary, in the dark energy-filled Universe~($\rho+p>0$),~that mass will increase progressively, leading to the gradual expansion of the Universe verifying the theoretical prediction about the infamous dark energy. Again,~the conservation equation for generalized Rastall gravity theory, given by Eqn.(\ref{7})~leads to the following expression:
\begin{equation}\label{ac.4}
\left(\rho+p\right)=-\left[\frac{1-3\epsilon\lambda\left(1+\omega(z)\right)}{3H(1-4\epsilon\lambda)}\right]\dot{\rho}~,   
\end{equation}
Hence, by applying the above Eqn.(\ref{ac.4}) into Eqn.(\ref{mass_rate}) and taking integration on both sides, we finally get our desired mass equation as follows:
\begin{equation}\label{mass_equation}
M=\frac{M_0}{1+\frac{4\pi M_0}{3}\mathop{\mathlarger{\mathlarger{\int}_{\rho_0}^{\rho}}}{\left(\lambda\mathcal{B}_0+\mathcal{B}_1\right)\left[\frac{1-3\epsilon\lambda\left(1+\omega(z)\right)}{(1-4\epsilon\lambda)}\right]\frac{d\rho}{H(z)}}}.     
\end{equation}
Where, $\rho_{_{0}}=\rho_{_{DM0}}+\rho_{_{DE0}}$~,~is the present value of the energy density of our Universe~,~$\rho_{_{DM0}}$~and~$\rho_{_{DE0}}$ respectively being the energy densities of dark matter and dark energy and $M_0$ is the present value of the black hole mass.\\\\ 
Now, let us investigate the two dark energy parameterization models considered in this study one by one, together with constrained values of the involved parameter estimated in Table \ref{tab_1}~of Section \ref{S2}~and observe how these can affect the mass accretion process around the black hole in the context of generalized Rastall gravity.
\begin{itemize}
    \item {\bf{Model-1}~:}\label{SS3B1}
As {\bf{Model-1}}~,~we consider the CBDRM dark energy parameterization, and its detailed description is given in Section \ref{SS1A}.~So, further, let us observe the mass accretion process for this particular parameterization and its influence over the mass of the black hole for the values of the involved parameters constrained through the MCMC technique in Section \ref{S2} in the context of our considered generalized Rastall gravity theory.\\ 
In this connection, the required mass equation due to CBDRM parametrization  of dark energy is given in the following form:
\begin{equation}\label{mass_equation_1}
M=\frac{M_0}{1+\frac{4\pi M_0}{3}\mathop{\mathlarger{\mathlarger{\int}_{\rho_0}^{\rho}}}{\left(\lambda\mathcal{B}_0+\mathcal{B}_1\right)\left[\frac{1-3\epsilon\lambda\left(1+\omega(z)\right)}{(1-4\epsilon\lambda)}\right]\frac{d\rho}{H(z)}}} ,
\end{equation}
Where,~$\omega(z)$~is the equation of state of CBDRM parametrization~,~given by Eqn.(\ref{EOS1})~,~$H(z)$~is the corresponding Hubble parameter associated to CBDRM parametrization~,~given by Eqn.(\ref{HPE1})~and~$\rho$~is the energy density of CBDRM parametrization~,~given as follows:
\begin{equation}\label{total_density_DE_1}
\rho=\frac{3H^2_0}{8\pi G}\left[~\Omega_{_{m0}}\left(1+z\right)^{3\left(\frac{R-4\epsilon\delta}{R-3\epsilon\delta}\right)}+~\Omega_{_{DE0}}~(1+z)^{3(1+\omega_0)}(2+z)^{3\omega_{1}}\right].
\end{equation}\\
To analyze the changes in the mass of the black hole with the evolution of the Universe, we plot the mass vs. redshift graph for Model-1 in Fig.\ref{fig:my_label1}~.~From Fig.\ref{fig:my_label1}~,~we can easily conclude that accretion of the CBDRM parameterization of dark energy will contribute to the enhancement of the black hole mass. 
\begin{figure*}
    \centering
    \includegraphics[width=0.5\linewidth]{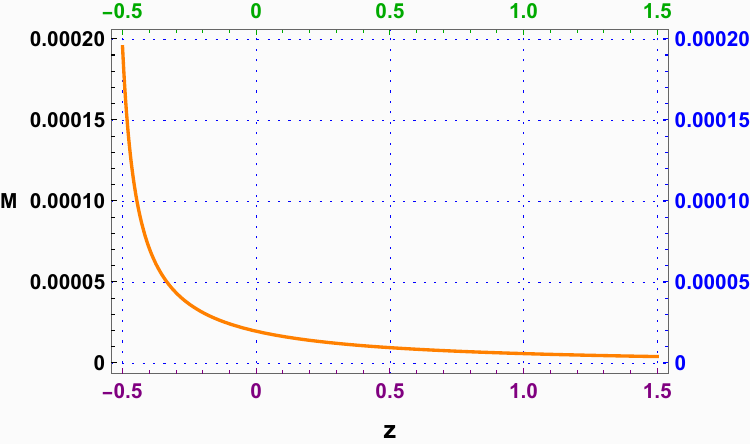}
    \caption{Changes in the mass of the black hole with redshift for Model-1~(CBDRM parametrization).}
\label{fig:my_label1}
\end{figure*}
\end{itemize}
\begin{itemize}
    \item {\bf{Model-2}~:}\label{SS3B2}
As {\bf{Model-2}}~,~we consider the dark energy parameterization of CDMMA~,~and a detailed discussion of this parameterization is given in Section \ref{SS1B}. Now, let us scrutinize the mass accretion process for this particular parameterization and its influence over the mass of the black hole for the values of the involved parameters constrained through the MCMC technique in Section \ref{S2} in the context of our considered generalized Rastall gravity theory.\\\\
In this connection, the required mass equation due to CDMMA parametrization of dark energy is given as follows:
\begin{equation}\label{mass_equation_2}
M=\frac{M_0}{1+\frac{4\pi M_0}{3}\mathop{\mathlarger{\mathlarger{\int}_{\rho_0}^{\rho}}}{\left(\lambda\mathcal{B}_0+\mathcal{B}_1\right)\left[\frac{1-3\epsilon\lambda\left(1+\omega(z)\right)}{(1-4\epsilon\lambda)}\right]\frac{d\rho}{H(z)}}} ,    
\end{equation}\\
Where, $\omega(z)$~is the equation of state of CDMMA parametrization~,~given by Eqn.(\ref{EOS2})~,~$H(z)$~is the corresponding Hubble parameter associated to CDMMA parametrization~,~ given by Eqn.(\ref{HPE2})~and~$\rho$~is the energy density of CDMMA parametrization~,~given as follows:
\begin{widetext}
\begin{equation}\label{total_density_DE_2}
\rho=\frac{3H^2_0}{8\pi G}\left[~\Omega_{_{m0}}\left(1+z\right)^{3\left(\frac{R-4\epsilon\delta}{R-3\epsilon\delta}\right)}+~\Omega_{_{DE0}}~(1+z)^{3\left(1+\omega_{0}+\frac{\alpha}{\omega_{1}}\right)}\left[\omega_{1}+\omega_{2}(1+z)^{\beta}\right]^{\frac{3(\omega_{1}-\alpha \omega_{2})}{\beta \omega_{1}\omega_{2}}}\right].
\end{equation}    
\end{widetext}
Now, to study the changes in the mass of the black hole with the evolution of the Universe, we plot the mass vs. redshift graph for Model-2 in Fig.\ref{fig:my_label2}~.~From Fig.\ref{fig:my_label2}~,~we can see that once again the black hole mass is increasing as a result of the accretion of the CDMMA parameterization of dark energy. 
\begin{figure*}
    \centering
    \includegraphics[width=0.5\linewidth]{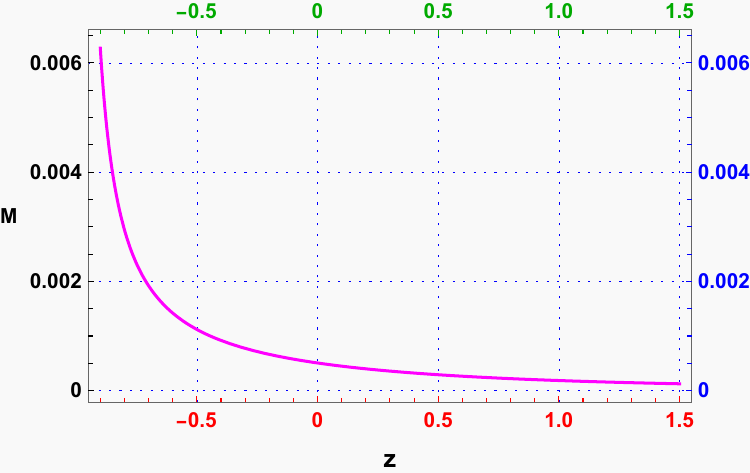}
    \caption{Changes in the mass of the black hole with redshift for Model-2~(CDMMA parametrization).}
\label{fig:my_label2}
\end{figure*}
\end{itemize} 
\begin{figure*}
\begin{subfigure}{0.32\textwidth}
\includegraphics[width=\linewidth]{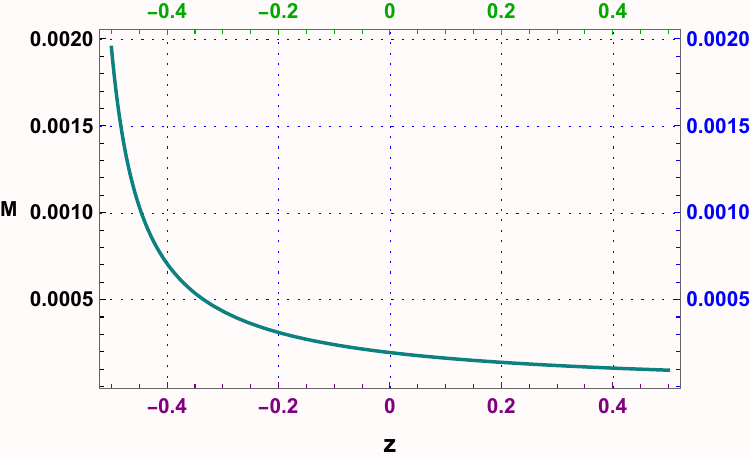}
    \subcaption{$M$ vs $z$ graph for Model-1~(CBDRM parametrization)}
    \label{fig:f 3}
\end{subfigure}
\hfill
\begin{subfigure}{0.32\textwidth}
\includegraphics[width=\linewidth]{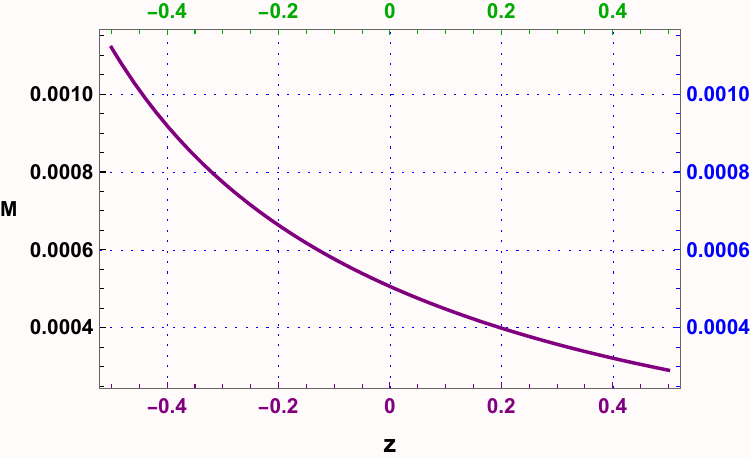}
    \subcaption{$M$ vs $z$ graph for Model-2~(CDMMA parametrization)}
    \label{fig:f 3}
\end{subfigure}
\hfill
\begin{subfigure}{0.32\textwidth}
\includegraphics[width=\linewidth]{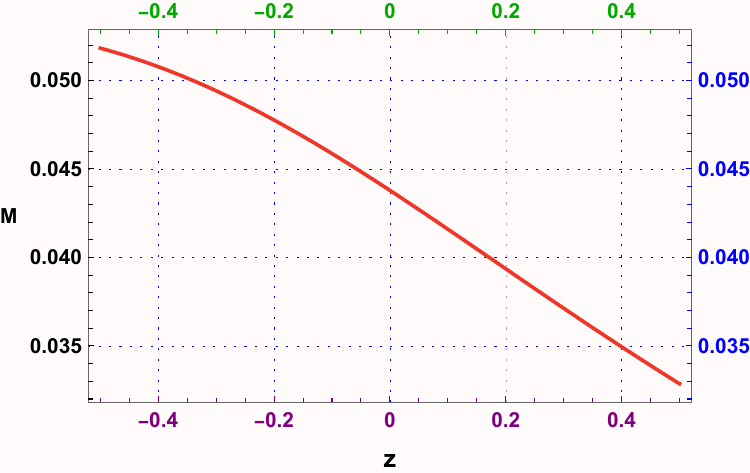}
    \subcaption{$M$ vs $z$ graph for $\Lambda$CDM Model of the Universe}
     \label{fig:f 3}
\end{subfigure}
\caption{}
\label{fig:my_label3} 
\end{figure*}
Thus, in the generalized Rastall gravity scenario, the accretion of dark energy parametrization models discussed above results in the rise of the black hole mass during the evolution of the Universe~,~which is compatible with the previously available studies related to the dark energy and the usage of the constrained values of parameters makes our results more practical~,~reliable~,~and accurate. Also, we draw a graphical comparison between these two models and the standard $\Lambda$CDM model of the Universe in Fig.\ref{fig:my_label3}~,~which gives us a fascinating view of the effects of dark energy on the black hole mass and verifies its true nature, the accelerated expansion of the Universe.\\
\section{Results~\&~Conclusions}\label{S4}
Black holes are one of the strangest creations of our Universe, which is why, after several years and so much research, many mysteries still hold it beyond complete human understanding. Mass is the most prominent and approachable property of a black hole, which can reveal many aspects of it. So, it is very convenient for researchers to study black holes in different gravities using their mass accretion process. Following that same approach, we have studied a non-singular black hole in the context of generalized Rastall gravity theory in this paper. Primarily, we have discussed the core concept and basic equations for the generalized Rastall gravity theory, which involve a modified version of the Friedmann equations and an explicit form of the generalized energy conservation equation. Considering the Universe to be a combination of dark matter and dark energy, we first obtained the energy density of dark matter. Then, we examined two dynamical dark energy equations of state parametrizations, labeled as \textbf{Model-1}~and~\textbf{Model-2}. As \textbf{Model-1}, we have chosen the CBDRM~parameterization and obtained its energy density equation together with the corresponding Hubble parameter in terms of density parameters in the framework of generalized Rastall gravity theory as our motive in this paper is to observe the effects of parameter constraining on the mass change of the black hole. The same has been repeated for the \textbf{Model-2}, where we have taken the CDMMA~parametrization. Then, we have constrained the required parameters using the Markov Chain Monte Carlo (MCMC) techniques in Section \ref{S2}~and compared the two parametrization models from the perspective of observational data, which indicated the supremacy of CDMMA parameterization above CBDRM parameterization whenever the best balance of fit quality and statistical significance is concerned.~Further, in Section \ref{S3}~,~firstly, we have discussed the non-singular black hole solution and its properties in the background of generalized Rastall gravity and then obtained their mass equation involving the density parameters for \textbf{Model-1}~and~\textbf{Model-2} respectively in terms of the redshift function. Now, to analyze the mass change of the black hole during the evolution of the Universe, we have plotted the mass vs. redshift graphs in Fig.\ref{fig:my_label1}~and~Fig.\ref{fig:my_label2}~respectively, which exhibited the accelerated expansion era of our Universe as per both the \textbf{Model-1}~,~and~the \textbf{Model-2}. In Fig.\ref{fig:my_label3}~,~we have displayed both of these models together with the standard $\Lambda$CDM model of the Universe side by side through their mass vs. redshift graphs just to visualize the difference in their mass-increasing patterns. All of these figures have been derived through the constrained values of the parameters, obtained in Table \ref{tab_1} of Section \ref{S2}~,~demonstrating the effects of parameter constraining on the changes of mass of the black hole in the environment of generalized Rastall gravity. In conclusion, this work predicts that, in the context of generalized Rastall gravity theory, the black hole mass will eventually increase for both the models of dark energy equation of state parameterization~:~CBDRM parametrization~and~CDMMA parametrization with the evolution of the Universe~,~which is abided by the true nature of mysterious dark energy and the consideration of constrained values of parameters makes our results more reliable and increases its accuracy further.\\\\     
\section*{ACKNOWLEDGEMENTS} 
PM is thankful to IIEST, Shibpur, India, for providing an Institute Fellowship (SRF).

\bibliographystyle{elsarticle-num}
\bibliography{Refer}

\end{document}